# Thermal conductivity and thermal rectification of nanoporous graphene: A molecular dynamics simulation

Farrokh Yousefi[a], Farhad Khoeini[a, *], Ali Rajabpour[b]

[a] Department of Physics, University of Zanjan, Zanjan, 45195-313, Iran
[b] Mechanical Engineering Department, Imam Khomeini International University, Qazvin, 34148–96818, Iran

**Abstract**

Using non-equilibrium molecular dynamics (NEMD) simulation, we study thermal properties of the so-called nanoporous graphene (NPG) sheet which contains a series of nanoporous in an ordered way and was synthesized recently (Science 360 (2018), 199). The dependence of thermal conductivity on sample size, edge chirality, and porosity concentration are investigated. Our results indicate that the thermal conductivity of NPG is about two orders smaller compared with of pristine graphene. Therefore this sheet can be used as a thermoelectric material. Also, the porosity concentration helps us to tune the thermal conductivity. Moreover, the results show that the thermal conductivity increases with growing sample length due to ballistic transport. On the other hand, along the armchair direction, the thermal conductivity is larger than zigzag direction. We also examined the thermal properties of the interface of NPG and graphene. The temperature drops significantly through the interface leading to the thermal resistance. The thermal resistance changes with imposed heat flux direction, and this difference cause significantly large thermal rectification factor, and heat current prefers one direction to another. Besides, to investigate those quantities fundamentally, we study the phonon density of states and scattering of them.

**Keywords:** Nanoporous graphene, Thermal conductivity, Thermal rectification, Molecular dynamics.

**Introduction**

Graphene[1], which is a nano-sized and two-dimensional (2D) sheet with honeycomb form, shows uniquely high mechanical properties[2], excellent thermal conductivity[3,4], optical or electronic properties[5]. Due to interesting properties of graphene, new classes of 2D materials were synthesized or predicted by the numerical method for different targets[6]. Although graphene has some outstanding properties, there are still some restrictions in practical usage. For example, the graphene band gap is near zero and behaves as a semimetallic material which restricts its usage in nanoelectronic and semiconductor industry. Therefore, the motivation rises to design and manufacture other materials without graphene limitations. This issue encouraged the people around the world to design and synthesis novel 2D nanosheets such as borophene[7], phospherene[8], Molybdenum disulfide[9], and carbon nitride[10]. As an example, recently a 2D semiconductor nanosheet so-called twin-graphene was predicted by first-principles calculations with 0.981eV band gap[11].

There is still some way to use awesome graphene properties in various situation. Graphene properties such as electronic, thermal, and optical show strongly tunable with impurity doping[12], defects[13], and mechanical straining[14]. Alongside these tuning methods, the lithography technique is applicable on graphene sheet to produce a various configuration that introduced recently[15]. The phonons trap in the nanomesh or graphene kirigami[16] that created by lithography and therefore thermal conductivity is reduced, or even band-gap is opened in graphene successfully. These structures can also be employed in other technologies such as water purification[17]. However, it is better to produce the nanoporous graphene in the chemical process, because the lithography is an additional action after synthesis the pristine graphene as well as it costs a lot of money and time.

Recently, Moreno et al.[18] synthesized and fabricated a novel 2D graphene sheet with nanoporosity (NPG) with a bottom-up method. This molecule shows inherent semiconducting character.[19] To employ the NPG in a practical situation, it needs to understand specifications and properties in a deep way. Thermal properties are the main issue in our current work that we study it in various conditions. Thermal conductivity of the NPG was obtained versus sample length, defect concentration, and two edge directions. Also, we investigated the thermal properties of the interface of hybrid NPG-graphene. Thermal properties of the interface between two types of materials have always been fascinating. According to previous works[20,21] on the interface between materials studied due to their importance and application in phononics. Moreover, the thermal diode or thermal rectifier is a system in which thermal resistance depends on the direction of heat current flow.[22]

**Computational method**

In the semiconductor and semimetals, the contribution of phonons in heat carriers is greater than the electrons that present in the system. According to the previous work[19] the NPG is a semiconductor so that we can ignore the electron contribution in the thermal conductivity[23]. In this work, all of NEMD simulations were performed using LAMMPS[24] package. The simulated system (Fig. 1a) is the NPG with a width of 10 nm, but the length can be different in various simulations. The thickness of the NPG was assumed 3.4 Å same as pristine graphene. Periodic boundary conditions were applied along in-plane (xy plane) directions while that the free boundary condition was considered for out-plane. To describe the interaction between carbon and hydrogen atoms in the NPG, it needs to choose an appropriate potential. According to the previous study by Bohayra *et al.*[19] we have employed AIREBO[25] Potential, which is a three-body potential. The AIREBO cutoff was set to 2.5 Å. The timestep assumed 1 fs in order to integrate the motion equation via the velocity Verlet algorithm. First, to relax structure and remove extra stress, the whole of the system was coupled using the Nosé-Hoover[26] barostat and thermostat at zero pressure and room temperature. After that, two narrow layers at the two ends of the sheet selected as fixed atoms so that they cannot move during simulation time. Two other regions near to the fixed atoms were specified as hot and cold baths. To generate a temperature gradient across the sheet, we used Nosé-Hoover thermostat (NVT) for the hot and the cold baths with a difference temperature of 40 K. Also, the NVE ensemble was applied to the region between baths. To calculate the temperature gradient, we divided the system into some slabs along with heat current direction with a width of 3 nm and the temperature of any slabs computed. Therefore, the slope of the temperature profile in that direction indicates the temperature gradient value. When applying the temperature gradient, the heat flux (transferred energy per (time×area)) reaches to steady state after 1.5 ns, in this case, the heat flux starts to fluctuate around the final value. The accumulative energy added into the hot slab or removed from the cold slab was saved to use in calculating the power of baths and the heat flux. Ensemble averaging done for the temperature gradient over 1 ns and used to calculate thermal conductivity using one-dimensional Fourier's low $\kappa = -\frac{j}{dT/dx}$ in which $j$ and $dT/dx$ are the heat flux and the temperature gradient.

On the other hand, we calculated thermal rectification factor. For this purpose, we built a hybrid system that includes the NPG and pristine graphene (Fig. 1b, c). In order to obtain thermal rectification factor, two simulations were performed on the NPG-graphene sheet. The system was imposed to a positive temperature gradient once and once again imposed to a negative temperature gradient. In both states, we are using the calculated heat flux to determine thermal rectification factor via $TR\% = \frac{j_n - j_p}{j_p} \times 100$, where $j_p$ and $j_n$ are the heat flux in positive and negative directions, respectively[27].

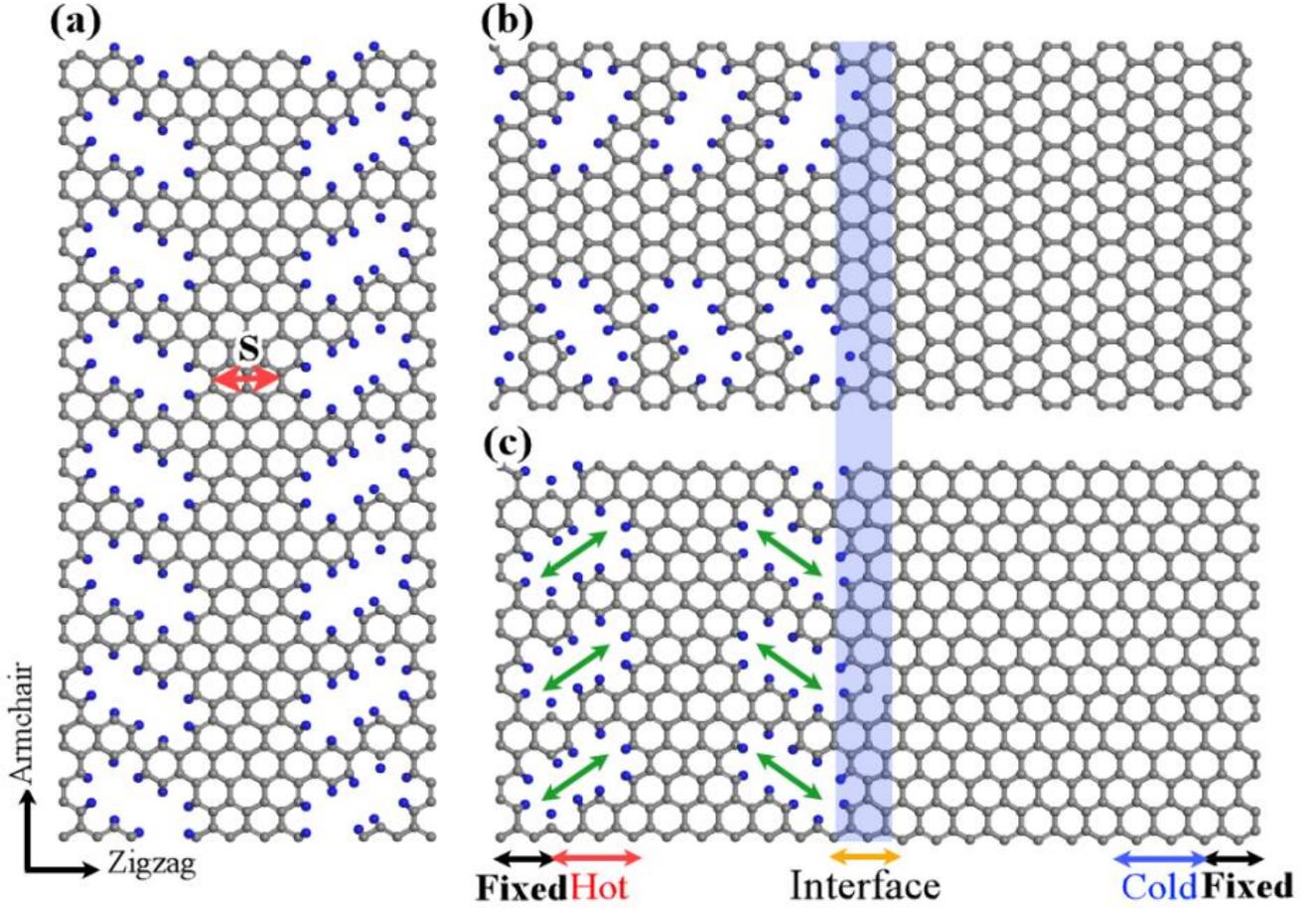

FIG. 1. (a) The schematic view of the NPG, (b) and (c) the schematic views of the hybrid of the NPG-graphene in the armchair and zigzag directions, respectively, which are used as a thermal rectifier. Hydrogen and carbon atoms colored with blue and gray, respectively. The parameter $S$ indicates the distance between the two consecutive defects that represents defect concentration in the NPG and the hybrid monolayer.

Also, phonon density of state (DOS) was obtained via calculating Fourier transformation of the velocity autocorrelation (indicated by < >) using the equation below[28],

$$P(\omega) = \sum_s \frac{m_s}{k_B T} \int_0^\infty e^{-i\omega t} <\vec{v}(0).\vec{v}(t)>_s dt \qquad (1)$$

where the summation and integration run over all atom types $s$ and the simulation time, respectively. Also, the quantities $m, \vec{v}, \omega$ are mass, velocity and phonon angular frequency.

## Results and discussion

All of the NEMD simulations were done to obtain thermal conductivity of the NPG, as well as the thermal rectification of the hybrid of the NGP-graphene. At first, as depicted in Fig. 2, the temperature profile of sheets with a length of 40 nm along the heat current direction is linear for the NPG with parameter $S=2$ where $S$ states the distance (the number of hexagonal) between two vertical lines of defects (see Fig. 1a). The temperature of each slab was calculated by equipartition theorem.

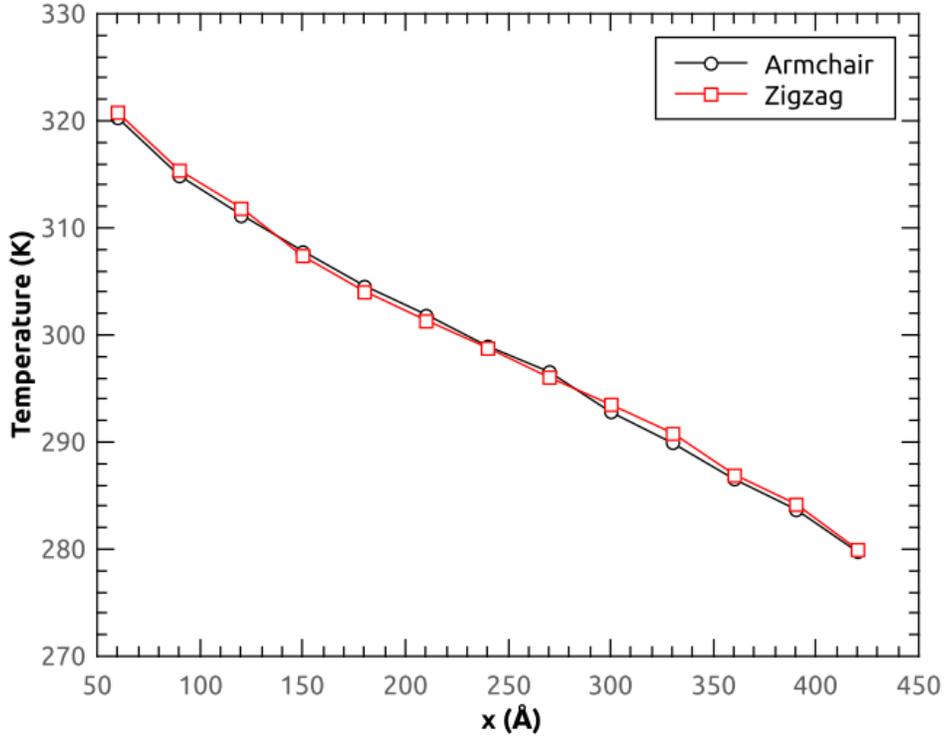

FIG. 2. Temperature profile for the NPG monolayer with $S=2$ along the armchair and zigzag directions.

Moreover, we have obtained accumulative energy that added into hot or removed from cold baths over simulation time. The data plotted in Fig. 3 for $S=2$ and pristine graphene in the armchair and zigzag directions. The slope of the curves ($\frac{dE}{dt}$) is the energy power (or heat current) provided by the hot bath thermostat. By dividing the energy power with the cross-section area of the sheet, the heat flux can be determined. As shown in the Fig. 3, the heat current that flows along the armchair direction of the NPG is greater than the zigzag direction [19].

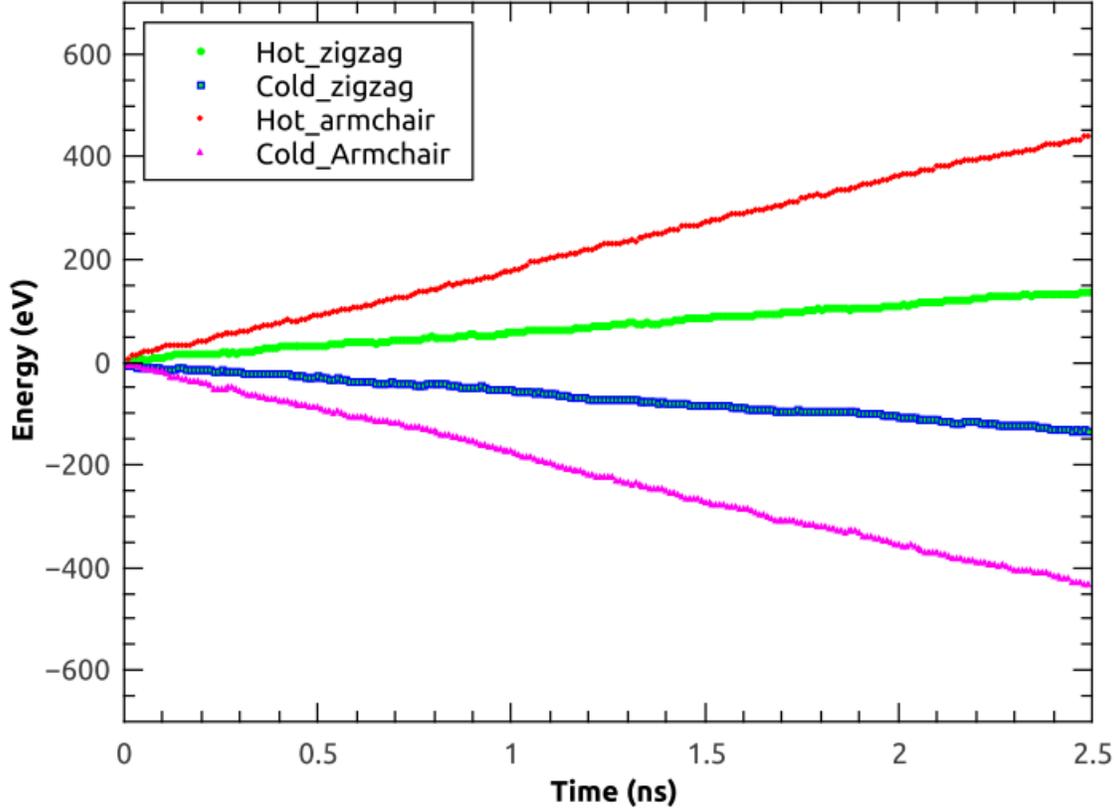

FIG. 3. Accumulate kinetic energy added into the hot and removed from the cold bath for two armchair and zigzag directions in the NPG.

Now, we will look more insightful description of thermal resistance that arises from the defects on the NPG nanosheet. According to Fig. 1a, there is an array of vertically oriented defects. When phonons (heat carriers) move toward to the defects, several phonons will be scattered. Therefore a few numbers of them will pass from defects. The thermal resistance decreases the efficiency of phonon transport. We have calculated thermal resistance in one of line defect by assuming that there is only one line defect in the nanosheet (see inset in Fig. 4). As defined, the thermal resistance is given by $R =$ $\frac{\Delta T_{interface}}{j} = 9.87 \times 10^{-11} m^2 K/W$, where $\Delta T_{interface}$ is the temperature drop in the interface. Suppose, there are some vertically oriented line defects in the NPG nanosheet which have thermal resistance $R$. The total thermal resistance[22] due to defects is $R_{total} = nR$, where $n$ is the number of line defects.

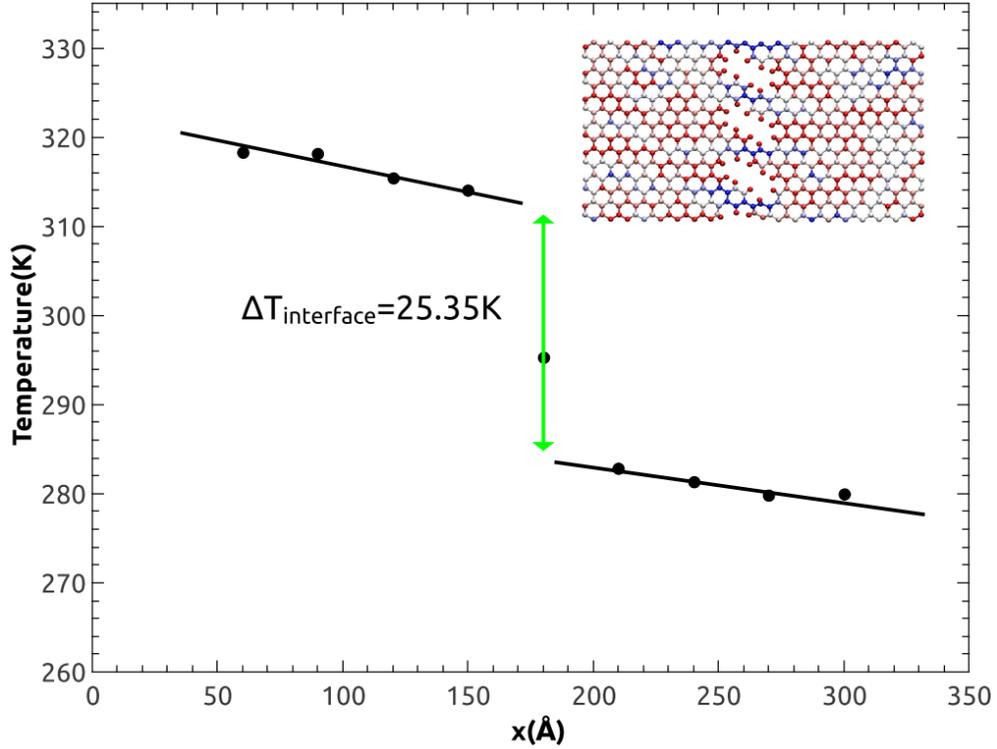

FIG. 4. Temperature drop in the vertically oriented defect clusters. Sample length is considered 30 nm, and the temperature difference between two ends sheet is 40 K. In subset, blue-colored atoms indicate that they conduct more heat current than others.

The length dependence of the thermal conductivity of the NPG was also studied. Since there is a limitation in the boundary of nanosheet due to fixed atoms, some phonons cannot be excited and therefore have no contribution to heat transport for small length. Therefore, in this work, we have considered a range of sample length between 30 nm and 90 nm to explore length effect at room temperature along with the armchair and zigzag directions (see Fig. 5). Thermal conductivity of the NPG monolayer along the armchair direction is greater than the zigzag one. This is due to the type of defects along with both directions. In the armchair direction, there is a narrow layer between two vertically consecutive defects which do not contain defects.

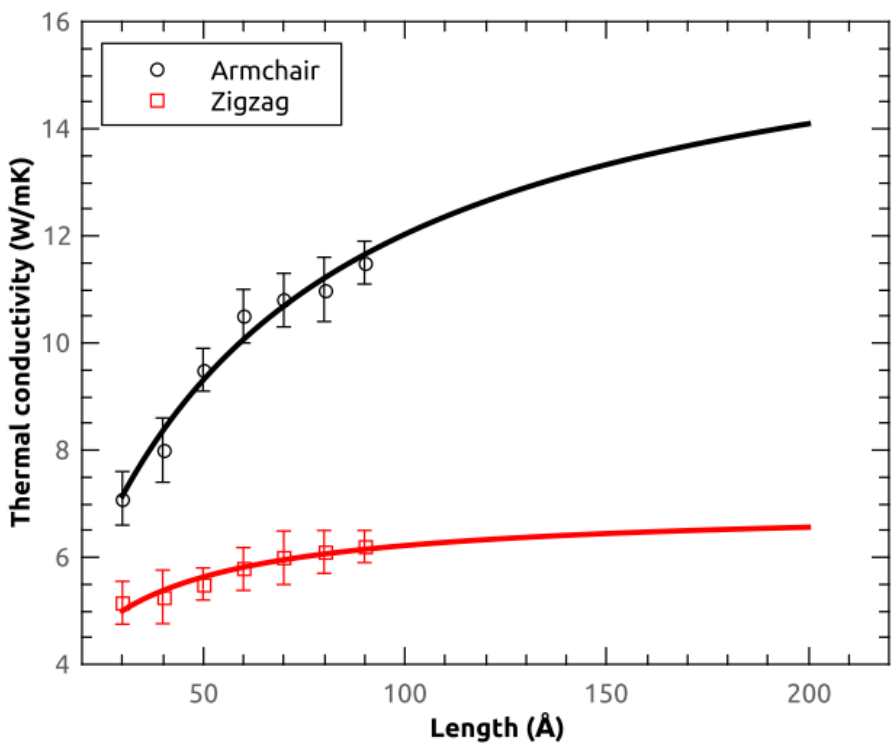

FIG. 5. Thermal conductivity of the NPG monolayer along with both directions versus sample length at room temperature. The fitting curve function is according to Eq. 2.

By increasing the sample length, two effects can occur. First, the more phonons will excite that lead to increase the heat conduction (positive effect). Second, the phonon-phonon and phonon-defects interactions will also lead to increase the thermal resistance and decreased thermal conductivity (negative effect). The mentioned effects compete to specify the behavior of the thermal conductivity for various samples of length. As explained by Felix *et al.*[29] there is a common way to extrapolate the NEMD results to larger sample length.

In this way, the thermal conductivity of a large sample is a function of sample length and given by,

$$\kappa_L^{-1} = \kappa_\infty^{-1}(1 + \lambda/L) \qquad (2)$$

where $\kappa_\infty$ and $\lambda$ is thermal conductivity of infinite sample length and the phonon mean free path (MFP), respectively. In the $L \ll \lambda$ regime (ballistic) we have $\kappa_L \propto L$, and thermal conductivity increases linearly with sample length. In this regime, the effect of increasing the number of exciting heat carrier in thermal conductivity dominate to negative effects caused by increasing defects. In the ballistic regime, the phonons move without scattering via defects. On the other hand, in the $L \gg \lambda$ regime (diffusive), the effect of sample length on thermal conductivity is weaker. By fitting Eq. 2 to thermal conductivity in Fig. 5, we have found $\kappa_\infty$ and $\lambda$ which reported in Table 1. For comparison, the thermal conductivity of infinite graphene ($709.2\ W/mK$) was calculated by this manner and the same potential in the previous work[30]. Also, this result is in agreement with the previous work[19]. As the last point of our study in this section, it is promising that the semiconductor NPG can be used as thermoelectric material because of its small thermal conductivity and get high performance for power generators[31].

Table 1. Thermal conductivity and the MFP for infinite length (*S*=2) obtained from fitting curve extrapolation according to Eq. 2.

| Direction | $\kappa_\infty$(W/mK) | $\lambda$(nm) |
|---|---|---|
| Armchair | 17.01 | 41.4 |
| Zigzag | 6.9 | 11.6 |

We have also examined the effect of the defect concentration (indicated as parameter *S* in Fig. 1a) on the NPG sheet. The parameter *S* changes from 2 to 7. The defect concentration decreases by increasing the value of *S*. As we expected, thermal conductivity should be increased for larger *S* due to reduced phonon scattering. As illustrated in Fig. 6, the NEMD results also show that the behavior of thermal conductivity along with both directions is incremental. The slope of thermal conductivity along the zigzag direction is insignificant because in this direction there is at least one sharp defect which does not allow pass heat flux enough. This explanation is not true for armchair direction since there is no sharp defect in that direction.

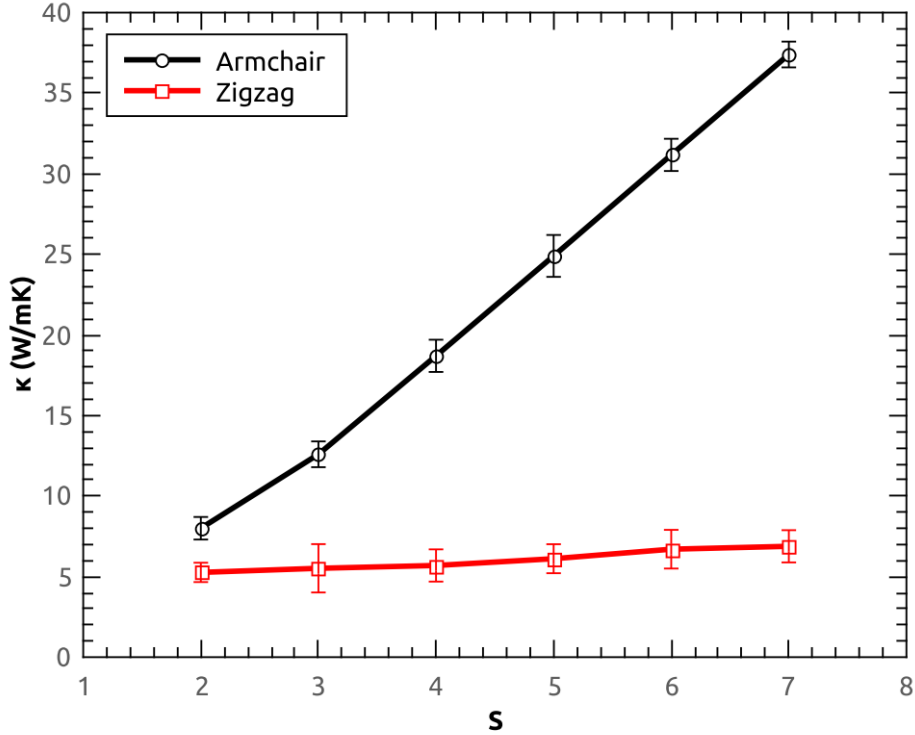

FIG. 6. Thermal conductivity of the NPG monolayer along both directions versus parameter $S$ for a length 50 nm at room temperature.

Hereafter, using the NEMD method the entire 15 ns simulation time, we study thermal rectification in the hybrid of NPG-graphene (see Fig. 1b, c). The NPG monolayer with parameter $S=2$ with armchair and zigzag directions was used in this combination. As mentioned above about methodology, once the system was imposed to a positive gradient and once again to a negative gradient. We denote the calculated values of the heat flux with $j_p$ and $j_n$ for positive and negative gradient, respectively. As shown in Fig. 7a, the temperature profile was plotted along with the zigzag direction with positive (left to right) and negative (right to the left) directions. The slope of the profile on both sides of the interface differs due to thermal resistance at the interface. The temperature gap in the interface is small.

Moreover, in Fig. 7b, the accumulative energy of hot and cold baths were plotted for both gradient directions. The slope of the accumulative energies divided by the cross-section of the sheet indicates $j_p$ and $j_n$. In order to extract thermal rectification factor, we have used the equation $TR\% = \frac{j_n - j_p}{j_p} \times 100$. The obtained $TR$ factors for armchair and zigzag directions are 4.66±0.5% and 6.01±0.6%, respectively. These $TR$ factors are small in analogy to previous works that were ~ 20%, thermal rectification for other systems[27].

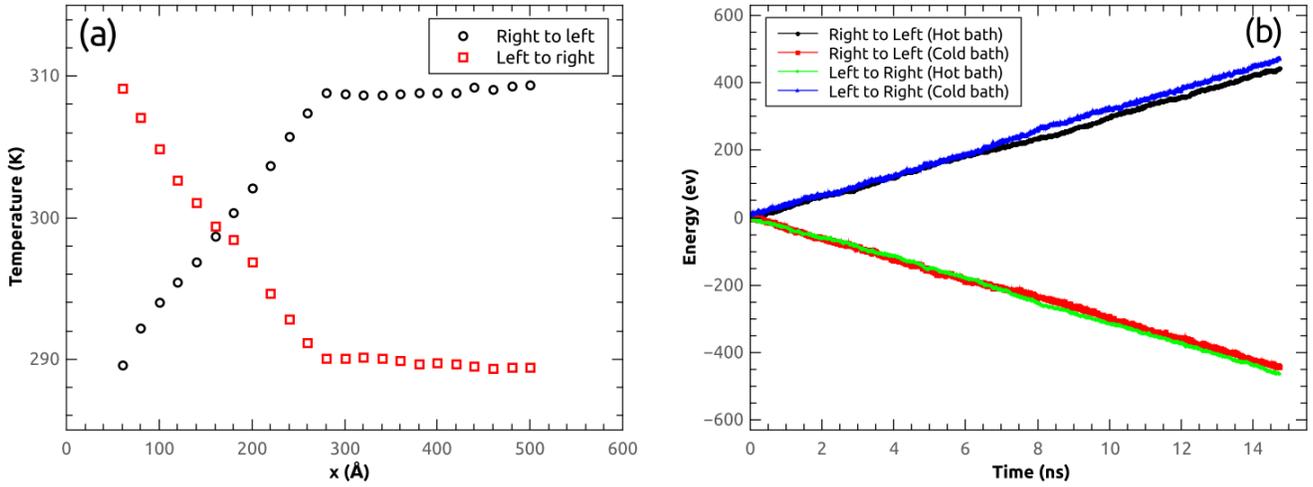

FIG. 7. (a) Temperature profile along with the zigzag direction. The positive and negative gradients were colored with red and black, respectively. The mean temperature is 300 K. (b) Accumulate energy that added to the hot bath and removed from the cold one for zigzag direction in the NPG-graphene.

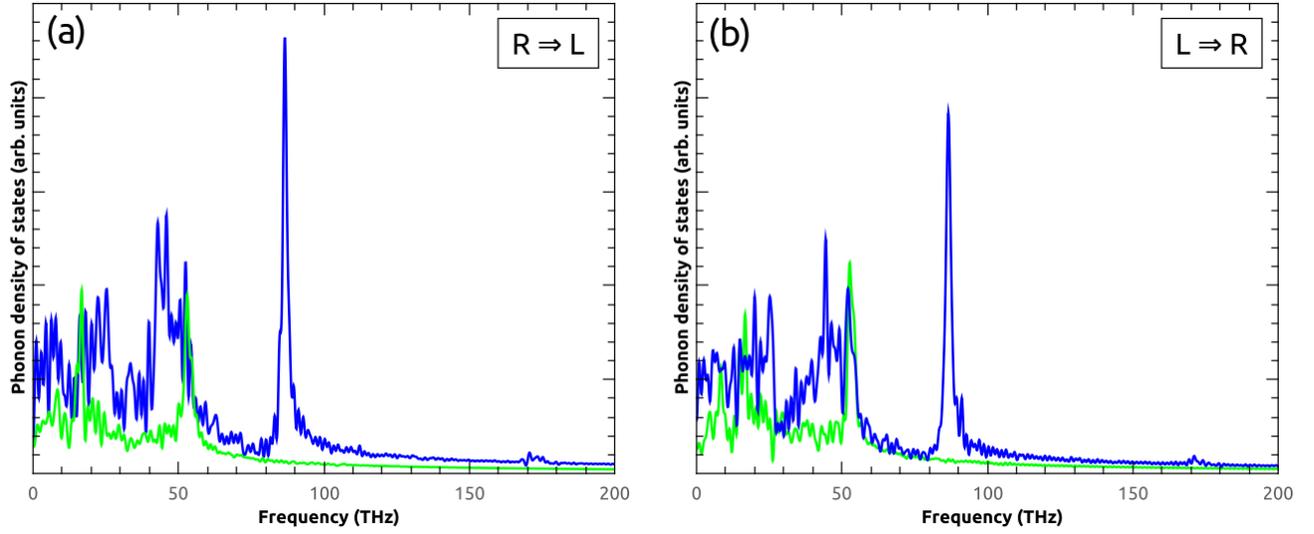

FIG. 8. Phonon density of states of two selected atoms on both sides of interface for the cases left to right and right to the left heat flux directions. Blue and green colored curves are for the NPG and graphene, respectively.

Another quantity that can help us to understand the fundamental mechanism of thermal rectification is the calculation of the phonon density of states (DOS). The DOS can be calculated according to Eq. 1, which uses the Fourier transform of the velocity autocorrelation. Two groups of atoms were selected on both sides of the interface, and the DOS was obtained. When two DOS spectra differ from each other, it shows that some phonons in the interface were scattered and could not pass from one side of the interface to another. As shown in Fig. 8, the DOS differs on both sides of the interface. Therefore, as we mentioned above, some of the phonons were scattered from the defect. As seen in Fig. 8, the blue-colored curve depicts the DOS of NPG monolayer, which includes carbon and hydrogen atoms. This curve shows that there are two peaks, one peak at around ~78 THz and other a small peak at ~170 THz due to the presence of hydrogen atoms in the system. These peaks arise from small mass of the hydrogen atoms, and therefore they can oscillate with high frequency.

**Conclusion**

In the present work, the NEMD simulations were carried out to explore thermal properties such as thermal conductivity and thermal rectification in the NPG monolayer and the hybrid of NPG-graphene for both armchair and zigzag directions. We investigated the length dependence of thermal conductivity and defect concentration. The NEMD results illustrated that thermal conductivity increases by increasing the sample length and also by decreasing the defect concentration. Thermal conductivity of the NPG was obtained small values $\leq 17.01$ and $\leq 6.9\ W/mK$ for any length of armchair and zigzag directions, respectively, which can be suitable material for thermoelectric devices. Moreover, the thermal rectification factors are also small and equal to 4.66±0.5% and 6.01±0.6% for armchair and zigzag directions, respectively. To describe the underlying mechanism of thermal rectification, the phonon density of states were calculated on both sides of the NPG-graphene interface. We illustrated that the obtained DOSs on both sides are mismatched, which is a significant factor of phonon scattering in the interface.

Email: khoeini@znu.ac.ir